\input{aipcheck}

\documentclass[
    ,final            % use final for the camera ready runs
  ]
  {aipproc}
\usepackage{cancel,amsmath,amssymb}
\usepackage{graphicx}
\usepackage{rotating,multirow}
\usepackage{wrapfig}
 \layoutstyle{6x9}

\begin{document}

\title{Diboson Physics at CDF}


\author{Alberto Annovi on behalf of the CDF Collaboration}{
  address={INFN - Laboratori Nazionali di Frascati, via E. Fermi 40, Frascati (RM), Italy}
}

\begin{abstract}
At the Fermilab Tevatron, the CDF detector is used to study diboson production in $p\bar{p}$ collisions at $\sqrt{s}$ = 1.96 TeV. We report recent diboson production measurements, limits on anomalous triple gauge couplings and latest results from semi-leptonic diboson searches.
\end{abstract}

\maketitle

%%%%%%%%%%%%%%%%%%%%%%%%%%%%%%%%%%%%%%%%%%%%
%% MAINMATTER
%%%%%%%%%%%%%%%%%%%%%%%%%%%%%%%%%%%%%%%%%%%%

\section{Introduction}
The study of diboson pair production is of high interest because it gives access to triple gauge couplings (TGC). It has the potential to unveil new physics that manifests as anomalous TGC (aTGC). 
Furthermore, it has a strong connection with the Higgs boson search, because diboson processes have finale states similar to those of Higgs production in association with a W or Z boson and are an important background for the Higgs search.

The CDF experiment has studied in great detail diboson processes generated from $p\bar{p}$ collisions at $\sqrt{s}$ = 1.96 TeV produced by the Fermilab's Tevatron.
We report on several recent diboson production measurements including latest results from semi-leptonic diboson searches.
For each channel the production cross-section is measured and compared with standard-model predictions.
In most case aTGC limits are set.

\section{Common diboson selections}
All analyses share common selection strategies that are summarized here.
The selection starts with either a W or a Z boson reconstructed in the leptonic decay mode. 
The first lepton is a trigger lepton with $p_T>20$~GeV. The second lepton is either a $\nu$ reconstructed as 
$\ensuremath{\cancel{E}_T}>20$~GeV or a second charged lepton with $p_T>10$~GeV.
When the second boson is a photon, it is reconstructed requiring $E_T >$ 7 GeV and $\Delta R(l,\gamma)>0.7$ where $l$ is any of leptons from the heavy boson.
When the second boson is either a W or a Z decaying leptonically the charged lepton(s) is reconstructed with $p_T>10$~GeV.
Finally, the second heavy boson is also reconstructed in the 2 jets final state with $E_T^{jet} >25$~GeV.

\section{W/Z+$\gamma$  production}
The process $p\bar{p} \rightarrow {\rm W}+\gamma$ is sensitive to the WW$\gamma$ coupling.
We measure the production cross-section using a data sample of 1.1~$\rm fb^{-1}$.
The angle between the photon and the charged lepton is used to separate the signal from the W+jets process that is the dominant background. The measured cross-section time branching ratio with $E_T(\gamma)>7$~GeV is $18.03 \pm 0.65\text{(stat.)} \pm 2.55\text{(sys.)} \pm 1.05\text{(lum.)}$ pb, which is consistent with the standard model prediction $19.3 \pm 1.4$ pb~\cite{Wgamma}.

The $p\bar{p} \rightarrow {\rm Z}+\gamma+X$ process is sensitive to the ZZ$\gamma$ coupling, which is 0 in the standard model. This process is similar to $\rm W + \gamma$ with the Z boson reconstructed in the $\rm e^+e^-$ or $\mu^+\mu^-$ channel.
In a sample of 2.0~$\rm fb^{-1}$, we measure the cross-section to be $4.6 \pm 0.2\text{(stat.)} \pm 0.3\text{(syst.)} \pm 0.3\text{(lum.)}$~pb that is in agreement with the SM prediction $4.5 \pm 0.3$~pb.
The photon could be emitted from initial or final state radiation. Any TGC effect would only be present in the initial state radiation (ISR) events. We measure the cross-section for ISR photons to be $1.2 \pm 0.1\text{(stat.)} \pm 0.2\text{(syst.)} \pm 0.1\text{(lum.)}$~pb . We extract aTGC limits: $|h^Z_3|< 0.083$, $|h^Z_4|< 0.0047$, $|h^\gamma_3|< 0.084$, $|h^\gamma_3|< 0.0047$~\cite{Zgamma}.   

\section{$\rm WW \rightarrow l^+l^-\nu\bar{\nu}$ cross section}
We measured the WW production cross section  in the two charged lepton ($e$ or $\mu$) and two neutrino  final state using  an integrated luminosity  of 3.6 fb$^{-1}$~\cite{WW}.
There are several backgrounds that produce this di-lepton signature that are rejected with specific cuts.
The first is $t\bar{t}$ which is suppressed with a jet veto (events at least one jet with $E_T>15$~GeV are rejected).
In order to reduce Drell-Yan, which otherwise would be O($10^5$) larger than the signal, we require $\ensuremath{\cancel{E}_T}>25$~GeV.
Finally, di-lepton events from heavy flavor processes are rejected with an invariant mass cut: $M_{ll}>16$~GeV.
After this selection the signal to noise ratio is nearly one to one.
The signal is separated from backgrounds using matrix element based likelihood ratios. 
The WW cross section is extracted using a binned maximum likelihood method which best fits $LR_{\text{WW}}$ signal and background shapes to data (see fig.~\ref{fig:WW_vadim}).
A total of 654 candidate events are observed with an expected background contribution of $320\pm47$ events. The measured WW cross section is \mbox{$12.1\pm0.9\text{(stat.)}^{+1.6}_{-1.4}\text{(syst.)}$}~pb. It is in good agreement with the Standard Model prediction and it represents the most precise measurement up to date. An updated version of this analysis sets limits on aTGC~\cite{WW_aTGC}.

\section{$\rm WZ \rightarrow l^+l^-l^{\pm}\nu$ cross section and aTGC limits}
In order to search for WZ diboson production, we look for events where both bosons decay leptonically. 
This analysis uses a data sample of 1.9 fb$^{-1}$ and requires 3 electrons or muons plus $\ensuremath{\cancel{E}_T}$ (missing transverse energy)~\cite{WZ}. We find a total of 25 events where \mbox{$21.6\pm2.6$} were expected. This gives a cross section of \mbox{$4.3^{+1.3}_{-1.0}\text{(stat.)}\pm0.2\text{(syst.)}\pm 0.3\text{(lum.)}$}~pb, to be compared with a theoretical cross section of $3.7\pm0.3$~pb. We use the $p_T$ of the Z to test the WWZ vertex. 
We consider the anomalous triple gauge couplings  $\lambda$, $\Delta g$, $\Delta \kappa$ as defined in \cite{WZTGC1} and \cite{WZTGC2} and implemented in MCFM. In the Standard Model all three of these couplings are zero. We set the limits for $\Lambda=2.0$ TeV:  $-0.13<\lambda<0.15$, $-0.15<\Delta g<0.24$ and $-0.82<\Delta \kappa<1.27$.

\section{$\rm ZZ \rightarrow l^+l^-l^+l^-$ and $\rm ZZ \rightarrow l^+l^-\nu\bar{\nu}$ cross section}
We search for the low cross-section $\rm p\bar{p} \rightarrow ZZ$ process with fully leptonic decays.
This analysis is based on a data sample corresponding to 1.9 fb$^{-1}$ of integrated luminosity~\cite{ZZ}. 
We look in two channels. In first one both Z's decay to electrons or muons. 
In the second channel one Z boson decays in two neutrinos and the other decays to electrons or muons. 
We find 3 $\text{ZZ}\rightarrow \ell\ell\ell\ell$ candidates with an expected background of $0.096^{+0.092}_{-0.063}$
that gives a significance of $1.2\sigma$. This is combined with the $\text{ZZ}\rightarrow\ell\ell \nu \nu$ channel which used a matrix element discriminator to separate WW and ZZ, which finds a signal with a significance of $4.2\sigma$. The combined significance is $4.4\sigma$. The measure cross-section is \mbox{$1.4^{+0.7}_{-0.6}\text{(stat.+syst.)}$ pb} that agrees with the Standard Model expectation of 1.4~pb at NLO. An updated analysis reaches a significance of $5.7\sigma$ with the four-leptons channel alone~\cite{ZZ_4lep}.

\section{$\rm WZ/ZZ \rightarrow l^+l^-jj$ aTGC limits}
The ZZ process along with the WZ process are searched also in the semi-leptonic final state using a data sample of 1.9~fb$^{-1}$.
The Z boson is reconstructed in the $\rm e^+e^-$ or $\mu^+\mu^-$ channel.
Then, the di-jet mass spectrum for the $p_T(Z)>210$~GeV region is fit to constrain potential contributions from anomalous couplings. 
No WZ/ZZ excess is found. We set limits on ZZZ and ZZ$\gamma$ couplings for $\Lambda=1.2$ TeV: $-0.12<f_4^{\text{Z}}<0.12$, $-0.13<f_4^{\gamma}<0.12$, $-0.10<f_5^{\text{Z}}<0.10$, $-0.11<f_5^{\gamma}<0.11$.
We set limits on WWZ coupling for $\Lambda=2.0$~TeV: $-0.20<\Delta g<0.29$, $-1.01<\Delta \kappa<1.27$ and $-0.16<\lambda<0.17$. All details are available in refs.~\cite{ZZTGC,WZTGC}.

\section{First observation of $\rm VV \rightarrow \nu\bar{\nu}q\bar{q}$}
Diboson production has been observed at the Tevatron in the lepton channels through the leptonic
decays of the electroweak gauge bosons. Doing the same thing with jets is much more difficult due to
the large background from V+jets ($\text{V}=\text{Z}$, W).
%Those events are interesting because the topology of this events is similar to events where a Higgs boson is produced in association with a W or a Z, allowing diboson measurements to provide an important step towards possible future measurement of Higgs production. 
We present here the first observation at hadron colliders of $p\bar{p} \rightarrow VV$ decaying into $\cancel{E_T}$ plus jets~\cite{MetJJ}.

\begin{figure}
\begin{minipage}{0.49\textwidth}
\includegraphics[width=1\textwidth,height=0.3\textheight]{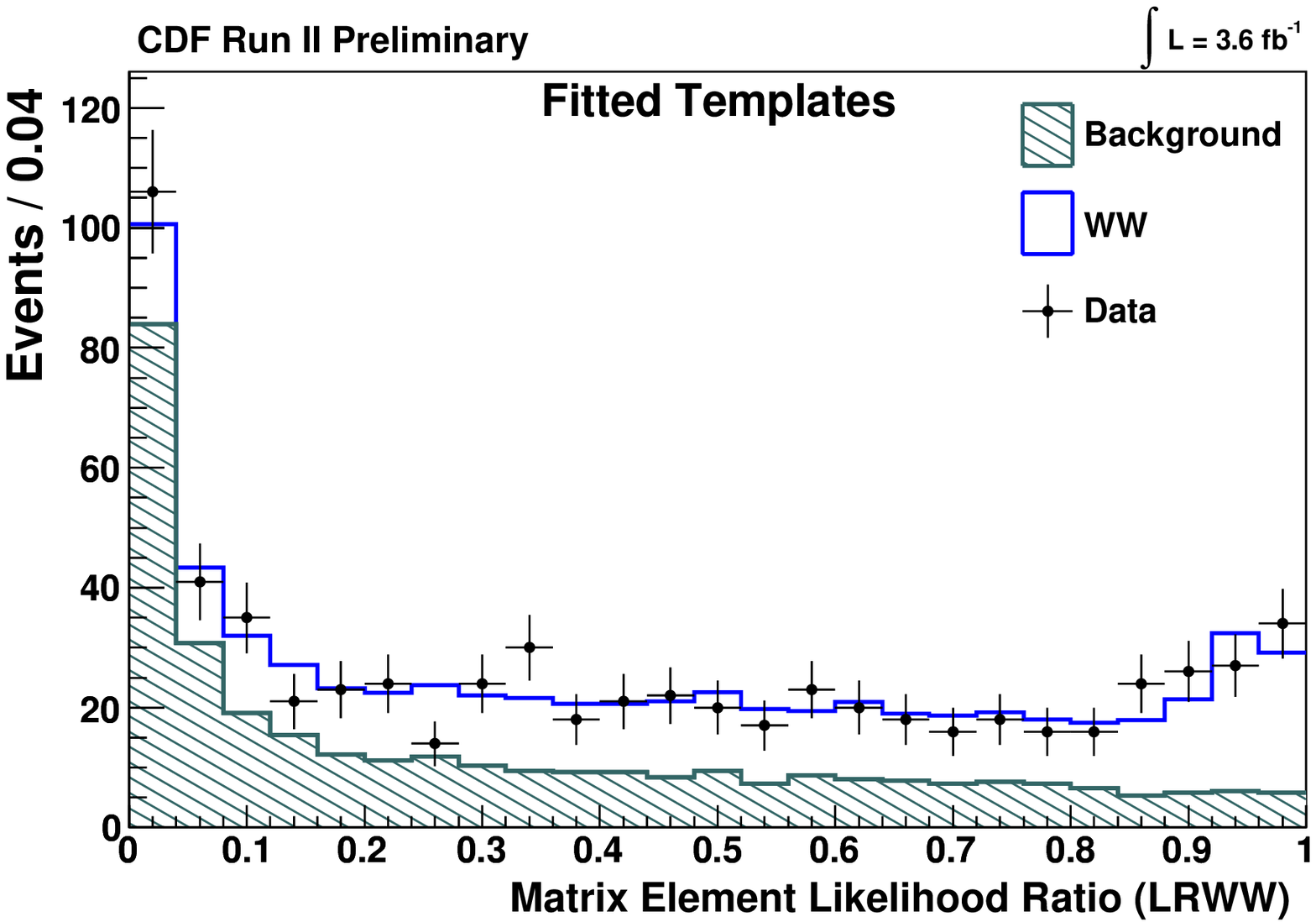}
\caption{\label{fig:WW_vadim}(left) Likelihood ratio used to extract the $\rm WW \rightarrow l^+l^-\nu\bar{\nu}$ signal. (right) Signal extraction fit for $\text{VV}\rightarrow\cancel{E}_T$+jets.}
\end{minipage}
\hspace{0.5cm}
\begin{minipage}{0.49\textwidth}
\includegraphics[width=1\textwidth,height=0.3\textheight]{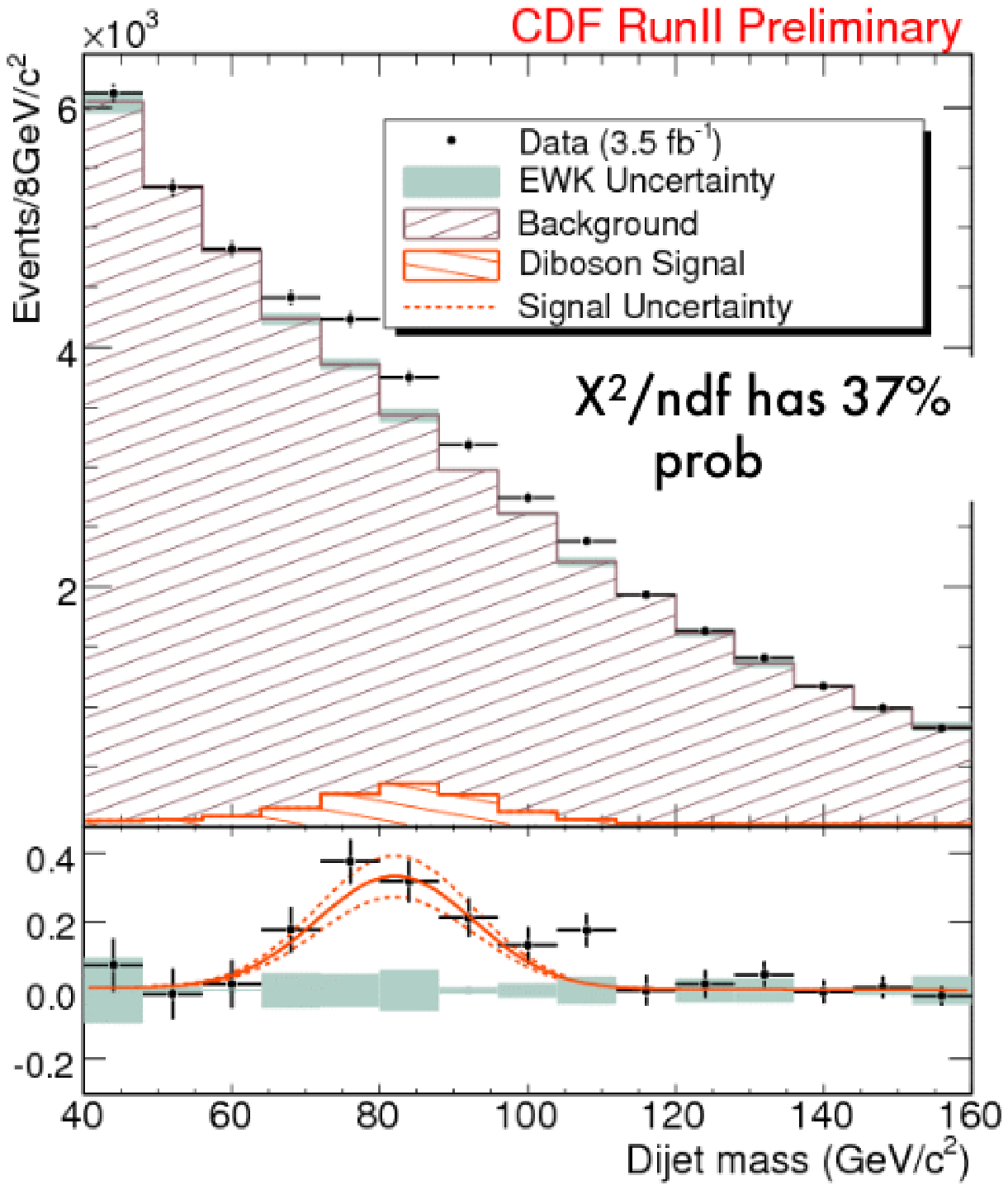}
\end{minipage}
\end{figure}

We measured the diboson cross-section using events with large $\cancel{E}_T$ ($>60$ GeV) and two jets with $E_T$ above 25 GeV. Due to limited energy resolution we cannot distinguish between WW, WZ and ZZ events so, what we measure is really a sum of all these processes in our selection window. No cut on number of leptons in the event is performed, therefore we are also sensitive to lepton decays of the gauge bosons. The QCD contribution, which is large in this channel, is heavily suppressed  through novel algorithms related to $\cancel{E}_T$ significance. We extract the signal from the background using the invariant mass distribution of the two jets in the event (see fig.~\ref{fig:WW_vadim}).
The extraction of the signal does not use the theoretical calculation of the V+jets integral cross section.
We observe $1516\pm239\text{(stat.)}\pm144\text{(syst.)}$ events that leads to a significance of $5.3\sigma$. 
We measure a cross section of \mbox{$\sigma (p\bar{p}\rightarrow VV+X)= 18.0\pm2.8\text{(stat.)}\pm 2.4\text{(syst.)}\pm 1.1\text{(lum.)}$ pb}, in good agreement with the Standard Model expectations.
Recent analyses also observed diboson production in the $\rm l\nu jj$ final state~\cite{lnujj}.

\section{Conclusions}
We have studied diboson production in $p\bar p $ collisions and have found it to be consistent with Standard Model predictions. We have set limits on a number of different couplings that do not appear in the Standard Model. 
We presented the first observation at hadron colliders of $p\bar{p} \rightarrow VV$ decaying into semi-leptonic final state.

%%%%%%%%%%%%%%%%%%%%%%%%%%%%%%%%%%%%%%%%%%%%%%%%
%% BACKMATTER
%%%%%%%%%%%%%%%%%%%%%%%%%%%%%%%%%%%%%%%%%%%%%%%%

%% \begin{theacknowledgments}
%% \end{theacknowledgments}



\bibliographystyle{aipproc}   % if natbib is available

%%\bibliography{sample}

\IfFileExists{\jobname.bbl}{}
 {\typeout{}
  \typeout{******************************************}
  \typeout{** Please run "bibtex \jobname" to optain}
  \typeout{** the bibliography and then re-run LaTeX}
  \typeout{** twice to fix the references!}
  \typeout{******************************************}
  \typeout{}
 }

\end{document}